\documentclass[prl,twocolumn,a4paper, superscriptaddress]{revtex4-1}
\usepackage[utf8]{inputenc}
\usepackage[pdftex]{graphicx}
\usepackage{color}

\newcommand{\mathbb}[1]{\mathcal{#1}}
\newcommand{\pp}{\mathbb{P}}
\usepackage{mathtools}
\DeclarePairedDelimiter{\evdel}{\langle}{\rangle}
\RequirePackage[normalem]{ulem} %DIF PREAMBLE
\RequirePackage{color}\definecolor{RED}{rgb}{1,0,0}\definecolor{BLUE}{rgb}{0,0,1} %DIF PREAMBLE
\providecommand{\DIFdelbegin}{} %DIF PREAMBLE
\providecommand{\DIFdelend}{} %DIF PREAMBLE
\begin{document}

%opening
\title{Random network models with geometric constraints}
\title{Spatial random network models for protein folding in 2D}
\title{Geometrically constrained random network models for protein folding in 2D}
\title{Geometrically constrained random network models & protein folding in 2D}
\title{Network Evolution, Geometric Constraints and Protein Folding}
\title{Random shortcuts, geometric constraints and protein folding}
\title{Network reorganization, geometric constraints and protein folding}
\title{Impact of geometric constraints on network reorganization}
\title{Impact of geometric constraints on network formation and protein folding}

\title{Spatial Constraints in Network Formation Dynamics and Applications to Protein Folding}
\title{Spatial Constraints in Network Structure Formation and Applications to Protein Folding}
\title{Spatial Constraints in Network Structure Formation and Applications to Protein Folding}
\title{Geometric Constraints in Network Structure Formation and Applications to Protein Folding}
\title{Impact of Spatial Contraints on the Dynamics of Network Structure Formation}

\title{Impact of Geometric Contraints on the Dynamics of Network Structure Formation}
\title{Spatially Fractal Networks: \\
Impact of Geometric Contraints on the Dynamics of Network Structure Formation}
\title{From Geometric to Network Constraints in Structure Formation Processes}
\title{Impact of Geometric Constraints on Spatial Network Formation
 and Applications to Protein Folding}
\title{Geometric Constraints and Scaling Laws in Spatial Network Formation}
\title{Scaling Laws of Geometrically Constrained Network Formation}
\title{Geometric Constraints Induce Scaling Laws in Spatial Networks}
\DIFdelbegin %DIFDELCMD < 

%DIFDELCMD < %%%
\DIFdelend \title{Geometric Constraints and Scaling Laws in Spatial Networks}
\title{Scaling Laws in Spatial Networks induced by Geometric Constraints}
\title{Scaling Laws in Spatial Network Formation}

\author{Nora Molkenthin}
 %\email[Electronic Address: ]{molkenthin@nld.ds.mpg.de}
 \affiliation{Network Dynamics, Max Planck Institute for Dynamics and Self-Organization (MPIDS), 37077 Göttingen, Germany}
 \affiliation{Department of Physics, Technical University of Darmstadt, 64289 Darmstadt, Germany} 
 \affiliation{Institute for Nonlinear Dynamics, Faculty of Physics, University of Göttingen, 37077 Göttingen, Germany}

 \author{Marc Timme}
 \affiliation{Network Dynamics, Max Planck Institute for Dynamics and Self-Organization (MPIDS), 37077 Göttingen, Germany}
\affiliation{Department of Physics, Technical University of Darmstadt, 64289 Darmstadt, Germany}
\DIFdelbegin %DIFDELCMD < 

%DIFDELCMD <   %%%
\DIFdelend \affiliation{Institute for Nonlinear Dynamics, Faculty of Physics, University of Göttingen, 37077 Göttingen, Germany}

\begin{abstract}
Geometric constraints impact the formation of a broad range of spatial networks, from amino acid chains folding to proteins structures to rearranging particle aggregates. How the network of interactions dynamically
self-organizes in such systems is far from fully understood. Here, we analyze a class of spatial network formation processes by introducing a mapping from geometric to graph-theoretic constraints.
Combining stochastic and mean field analyses yields an algebraic scaling law for the extent (graph diameter) of the resulting networks with system size, in contrast to logarithmic scaling known for networks without constraints. Intriguingly, the exponent falls between that of self-avoiding random walks and that of space filling arrangements, consistent with experimentally observed scaling (of the spatial radius of gyration) for protein tertiary structures.
\end{abstract}

\maketitle

Most networks forming in the real world are spatially extended and often geometrically constrained. Common examples include volume exclusion in the dynamics of polymers, chemical interactions in folding proteins and local electromagnetic forces in ferrofluidic aggregates \cite{Socci1994,Bundschuh1999,Taylor2010,Duncan2004,DeGennes1970,Koegel}. How geometric constraints impact the dynamic formation processes of spatial networks and thereby their function, is far from fully understood.

In many physical, chemical and biological systems, interaction structure and geometrical arrangement are equally important \cite{Barthelemy2011}, in particular for their dynamics. Key examples include proteins folding into their tertiary structures \cite{Scheraga2007,Saunders2013}. During the folding process, not only do amino acids interact with their neighbours along the chain but also with units that are far apart in the chain but close \emph{in space} \cite{Mirny2001,Shakhnovich2006}. On the level of abstract contact networks \cite{Dokholyan2002,Bagler2007a,DiPaola2013}, the process of protein folding can thus be considered as adding interaction links to a network, akin to percolation \cite{Stauffer1994, Nagler2011, Havlin2010}, but spatially transforming the network at the same time.

In this Letter, we demonstrate that geometric constraints induce algebraic scaling laws in the formation of spatial networks, suggesting self-similar (`fractal') structures.  We introduce a stochastic model that explicitly captures the essential impact of such geometric constraints on establishing spatial contacts and map them to constraints on graph-theoretic link additions. Combining probabilistic analysis with mean field calculations, we show that the extensions of the resulting networks exhibit an algebraic scaling law with system size.  In stark contrast, network formation processes without such constraints exhibit logarithmic scaling \cite{Newman2000} such that geometric constraints qualitatively change the nature of the scaling law. Intriguingly, the algebraic scaling law \emph{per se} as well as its exponent are consistent with the scaling of the experimentally observed spatial radius of gyration with the chain length of protein tertiary structures.

\begin{figure}[h]
        \centering
        \includegraphics[width=0.95\columnwidth]{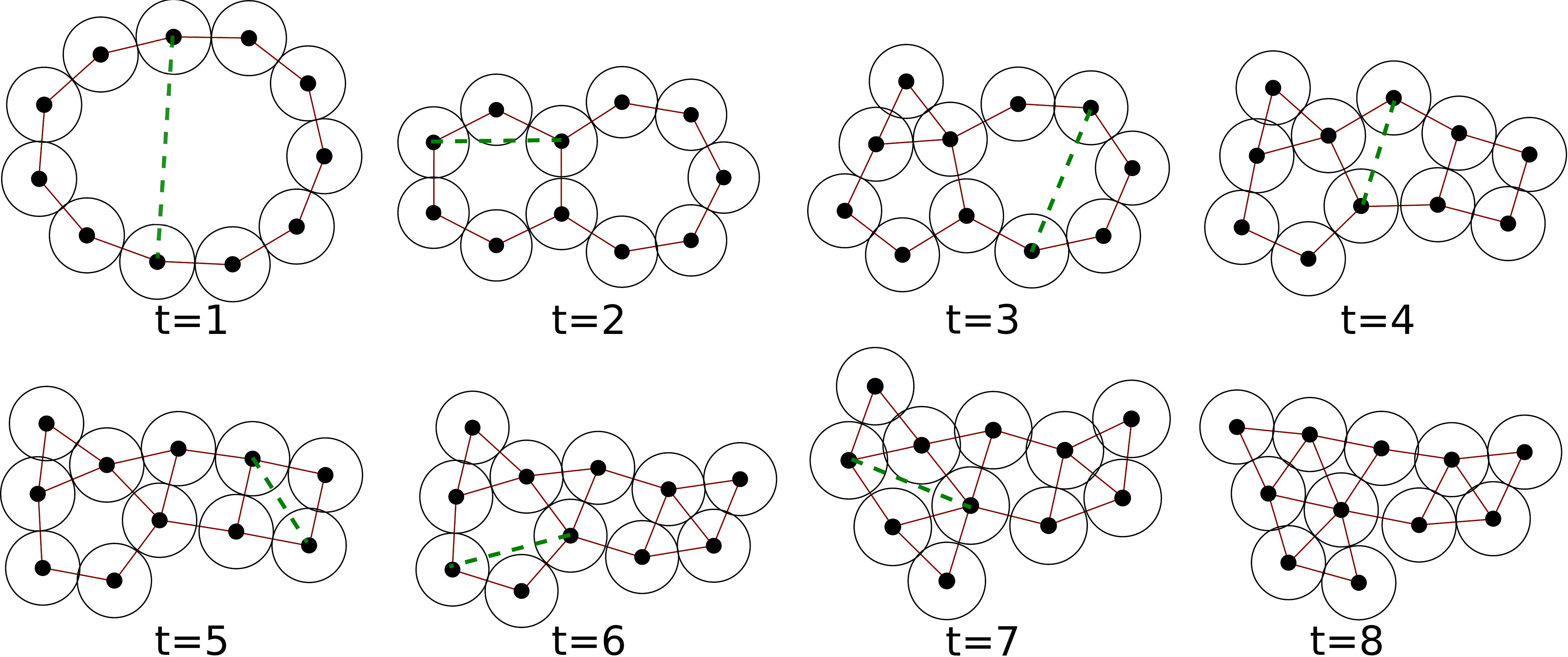}
        \caption{(color online) \textbf{Mapping spatial structure formation onto network formation.} Units coming into spatial contact (green dashed lines) induce additional links on the network level. The network becomes more and more compact as links add, in two dimensions yielding a subgraph of the triangular grid. For illustration, panels show networks of $N=11$  units for time steps $t \in \{1,2,\ldots,8\}$.}
        \label{fotos}
\end{figure}

\textit{Geometrically constrained network formation.}
To understand basic principles underlying geometrically constrained network formation dynamics consider an initial chain of identical, spatially extended units, each in contact with its nearest neighbor units (as all units are identical, this is a special case of a \textit{coin graph} \cite{Sachs1994}). For later analytic accessibility, we take the space to be two-dimensional and the chain to be closed to a single cycle such that initially the units are  indistinguishable. The latter does not change the scaling behaviour, because folding an open chain results in a collection of closed cycles, as we will see below. This chain represents the original aggregate such as an unfolded protein where the units are amino acids or an initial contact sequence of ferrofluidic particles. 

In a time-discrete network forming process (Fig.\,\ref{fotos}), the units randomly come into contact with each other under the geometric constraints that in each step (i) no two units overlap and (ii) units in contact at some point in time stay in contact. The chain thus non-locally deforms each time a new contact forms (Fig.\,\ref{fotos}). The sequence of connections models the emergence of pair-wise contacts between interacting units moving in space under the above constraints. In the model, new contacts keep forming until no additional contacts are consistent with the constraints. 
Thus, the resulting network is a collection of non-overlapping disks arranged to rigid triangles in two dimensional space (or spheres arranged to rigid hexagonal layers in three dimensional space). Checking whether this rigidity property can still be achieved for every newly established contact constitutes a non-local, computationally hard problem and is not simply feasible.

\begin{figure}[t]
        \centering
\includegraphics[width=0.75\columnwidth]{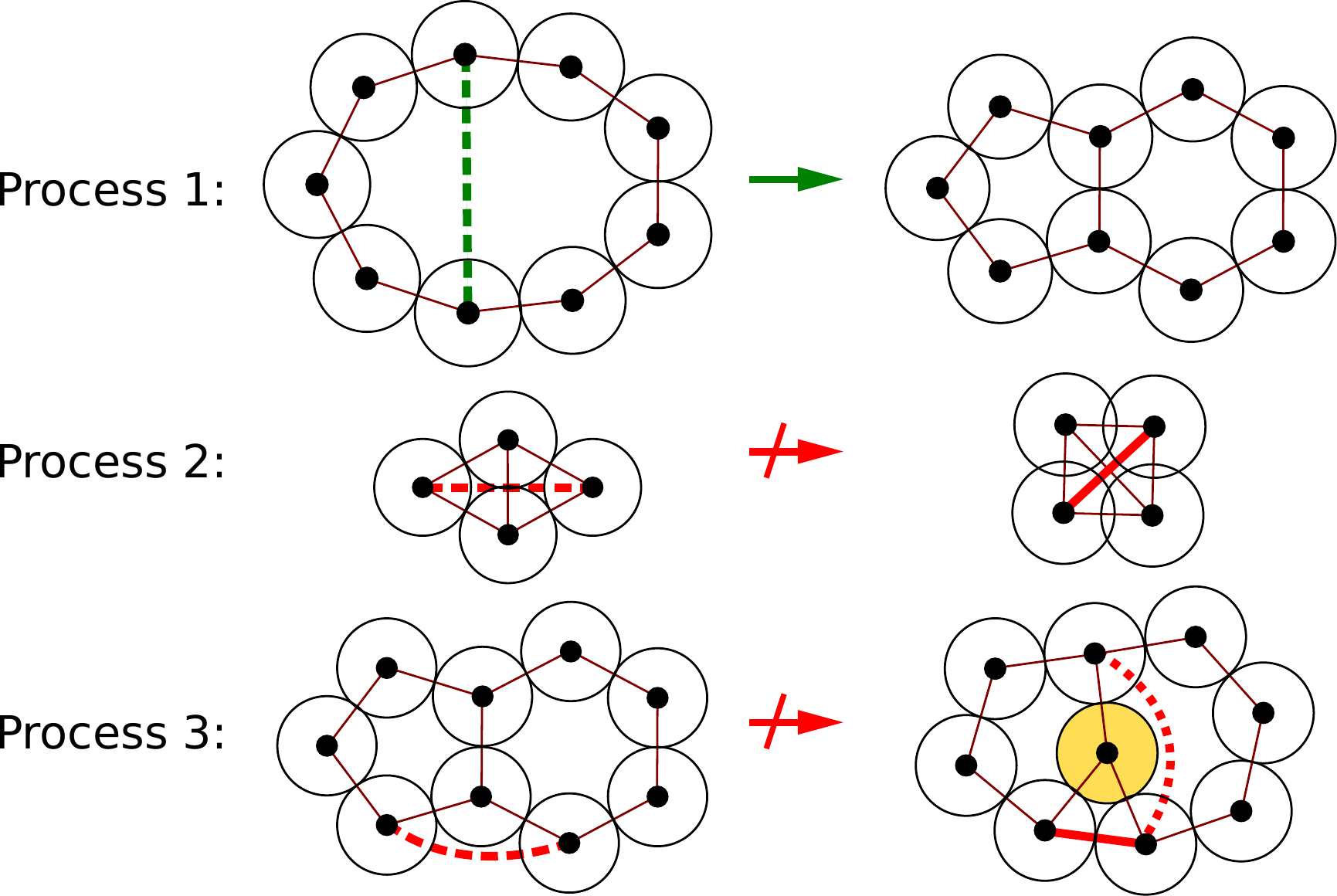} %was 0.55\columnwidth
        \caption{(color online) \textbf{Mapping constraints from spatial geometry to network topology.} Links in the contact graph form when two randomly chosen units come into spatial contact, subject to geometric constraints (a)-(d) specified in the text. Process 1: adding a link (green dashed line) is allowed because all conditions (a)-(d) are satisfied. Process 2: adding a link  (red dashed line) is forbidden due to condition (a) to avoid overlapping units. Process 3: adding a link on the outer face is forbidden due to condition (b) to avoid the possibility that units (here the one shaded yellow) may with later links (red dotted line) be enclosed by less than six other units during a subsequent step (e.g., red dotted line)}
        \label{constraints}
\end{figure} 
To analytically access the problem, we first map the spatial contact process with geometric constraints to a link addition process of network formation, with constraints on changes in the network topology only  (Fig.~\ref{constraints}). The map yields an approximate ensemble of networks that represent the spatial structure formation process. 
The topological constraints in the network model become:
(a) Links can only form between two units that are part of the same face of the graph (region enclosed by a cycle in the network). This ensures that geometric constraint (i) is not violated by links crossing. (b) Links do not form across the outer face. This ensures that no unit can be enclosed by less than six other units (which is geometrically impossible) such that (i) stays satisfiable. (c) The maximum degree of each unit is six. This also ensures that (i) is not violated by forcing more than six units around one given unit. (d) Once connected by a link, pairs of units do not disconnect, representing geometric constraint (ii). \\
\begin{figure}[h]
        \centering
	\includegraphics[width=\columnwidth]{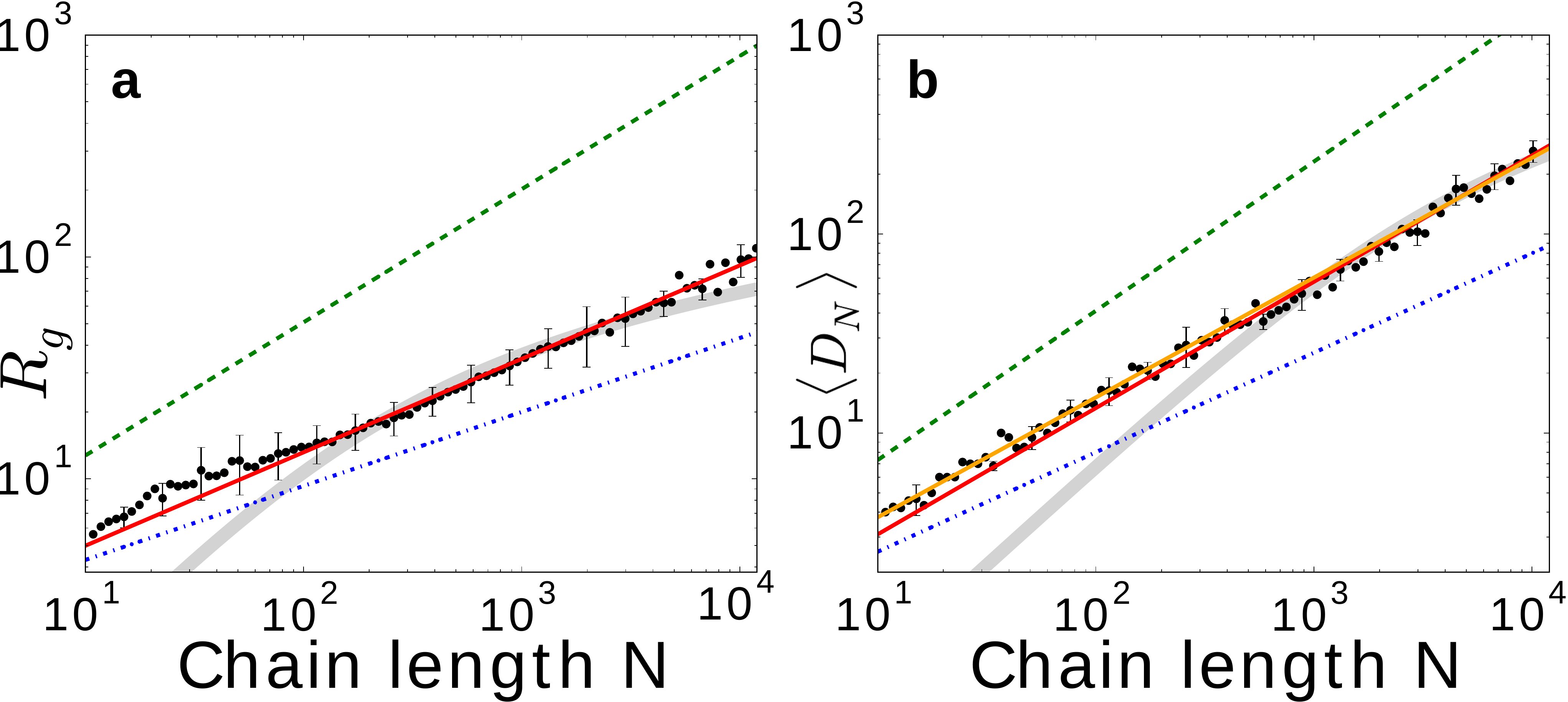}
        \caption{(color online) \textbf{Algebraic scaling laws in spatial network formation.} (a) Scaling of chain lengths of experimentally analyzed proteins vs. their radius of gyration (Eq.\,\ref{eq:RG}) (37162 data points from \cite{Hong2013} log-binned, with error bars indicating standard deviations). Best fits suggest algebraic (red) rather than logarithmic (gray) scaling \footnote{Logarithmic fitting by $R_g = a \ln(b N +c) - a \ln(c) $ ensuring that $\lim_{N\rightarrow 0} R_g(N)=0$}
        (b) Algebraic scaling of graph diameter $D_\textsf{final}(N)$ as derived in this Letter (orange line), plotted vs. the chain lengths $N$. Black dots indicate 450 stochastic realizations of network formation processes (uniformly sampled on a logarithmic chain length scale, binned and evaluated as in (a)) indicating the diameter of the original graph with best algebraic fit (red line). The algebraic scaling law with the (inverse) scaling dimension $\nu$ lying between that of self avoiding random walk $\nu_\textsf{RW}$ (green dashed lines) and that of space filling aggregates (blue  dotted lines) is consistent with biological data but inconsistent with logarithmic scaling as expected from network formation processes without geometric constraints.} 
        \label{num}
\end{figure}
\\
\\
\\
\\
\\
\textit{Spatial scaling of the network.} The spatial extension of an aggregate is often measured by its radius of gyration  
\begin{equation}
 R_\mathsf{g}=N^{-1}\Big(\frac{1}{2}\sum_{i,j}(r_i-r_j)^2 \Big)^{1/2} \sim N^\nu \, ,
 \label{eq:RG}
\end{equation}
quantifying the average distance between any pair out of $N$ units. Here,  $r_i$ is the spatial position of unit $i \in \{1,\ldots,N\}$ and $\frac{1}{\nu}$ is the scaling dimension. 
Real three-dimensional protein structures indeed exhibit an algebraic scaling law (Fig.\,\ref{num}a) with an exponent $\nu \approx 0.42 \pm 0.04$ above a lower bound $\nu_\textsf{SF}=1/3$ implied by compact space filling aggregates \cite{Hofmann2012,Rose2006,Danielsson2010,Hong2013} and below an upper bound $\nu_{\textsf{RW}}=3/5$ resulting from self-avoiding random walks in three dimensions without further restrictions \cite{Falck2003,Moore1978,Grosskinsky2002}, together yielding:
\begin{equation}
\nu_\textsf{SF} < \nu  < \nu_\textsf{RW} .
\label{eq:exponentBounds}
\end{equation}  
For spatially embedded networks where each unit occupies space of the same order of magnitude we expect the diameter $D$ to increase linearly with spatial extension.
Direct numerical simulations of the model processes for various system sizes indicate an algebraic scaling law
 \begin{equation}
D_\textsf{final} \sim N^\nu .
 \label{eq.Dscaling}
\end{equation}
as found for biological protein tertiary structures, see Fig.\,\ref{num}. Specifically, the obtained scaling exponent $\nu \approx 0.62 \pm 0.04$ moreover satisfies the same types of upper and lower bounds (Eq.\,\ref{eq:exponentBounds}) as experiments on proteins suggest, between space filling configurations (in two dimensions $\nu_\textsf{SF} = 1/2$) and that of self-avoiding random walks ($\nu_\textsf{RW} = 3/4$). 

\begin{figure}[h]
        \centering
	\includegraphics[width=0.5\columnwidth]{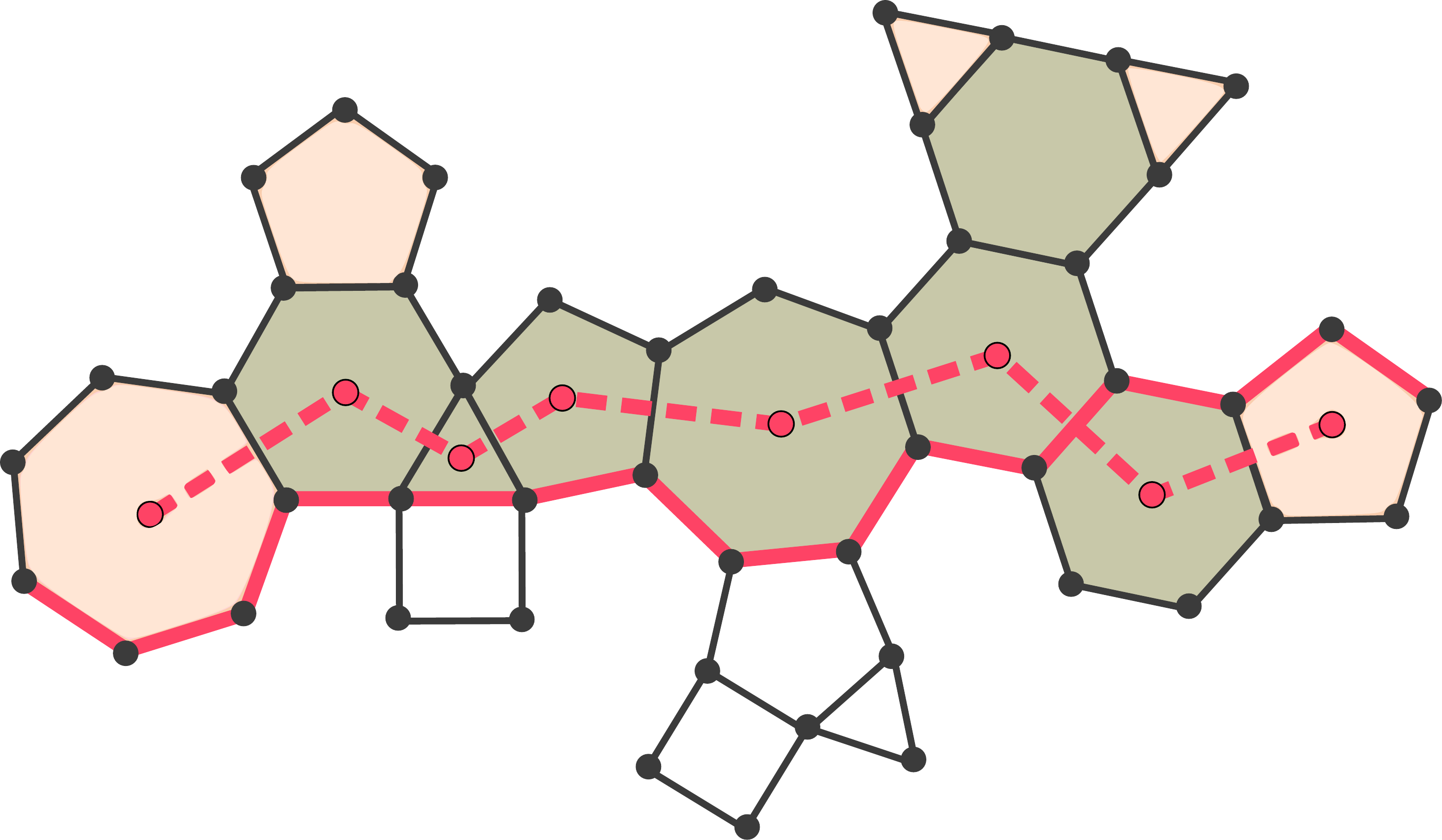}
        \caption{(color online) \textbf{Diameter path, diameter graph and end cycles.} A diameter path is a sequence of cycles of maximum length (here $D_t=7$, indicated by the dashed red line. For large graphs with defined average cycle length, $D_t$ is proportional to the diameter of the original graph (black dots, black solid lines, pink solid lines indicate diameter). The \textit{diameter graph} is the union of all such diameter paths (all shaded regions).  $V_t$  denotes  the number of cycles on the diameter graph (here $ V_t=12$) and $E_t$ the  number of end cycles (with only one neighbour) on any diameter path (here $E_t= 5$, shaded light rose).}
        \label{dual}
\end{figure}

\textit{Network formation integrating constraints.}
To understand the emergence of this scaling law and estimate its exponent, we mathematically analyze the network formation in the simplified network model with graph-theoretic constraints (a)-(d) inherited from the geometric ones (i) and (ii).

Consider at time $t=1$ an initial graph consisting of one  cycle of $N$ units that evolves in a process in discrete time $t\in\{1,2,\ldots \}$, with exactly one link adding at a time. Each new link divides one cycle into two smaller cycles. Such a process exclusively generates networks that are planar graphs consisting of cycles.

How does the above scaling emerge? How do the constraints impact the structure formation process on the network level? The graph-theoretical diameter of the dual graph of a given network serves as a natural quantity measuring the networks' extension. The vertices of the dual of a graph are defined by the faces of the cycles of the original, with two vertices connected if the two cycles they result from are neighboring, that is, share an edge in the original graph. At time $t$, the diameter $D_t$ of the dual therefore equals the length of (one of) its longest paths, representing a longest sequence of neighboring cycles in the original graph. We call such a sequence a \emph{diameter path}. The union of all diameter paths (all sequences of cycles of the same (largest) length) in the original graph is called the \emph{diameter graph}.

For small times $t$, the cycles are typically of different lengths, for larger times become similar and eventually all become triangles. Thus, for sufficiently large times $t$, the diameter of the network is proportional to that of the dual (Fig.\,\ref{dual}). We thus take a mean field view and simply talk about the \emph{diameter}, also when analyzing the scaling of the  the diameter of the dual. Since no two cycles share more than one link, and no unit of the original network becomes enclosed in any path (due to condition (b)), the resulting dual graph stays a tree at all times. The diameter graph thus is  the union of all paths of cycles of length $D_t$.
We note that  the total number of cycles present at that time $t$ equals $t$. 

\begin{figure}[t]
        \centering
        \includegraphics[width=0.8\columnwidth]{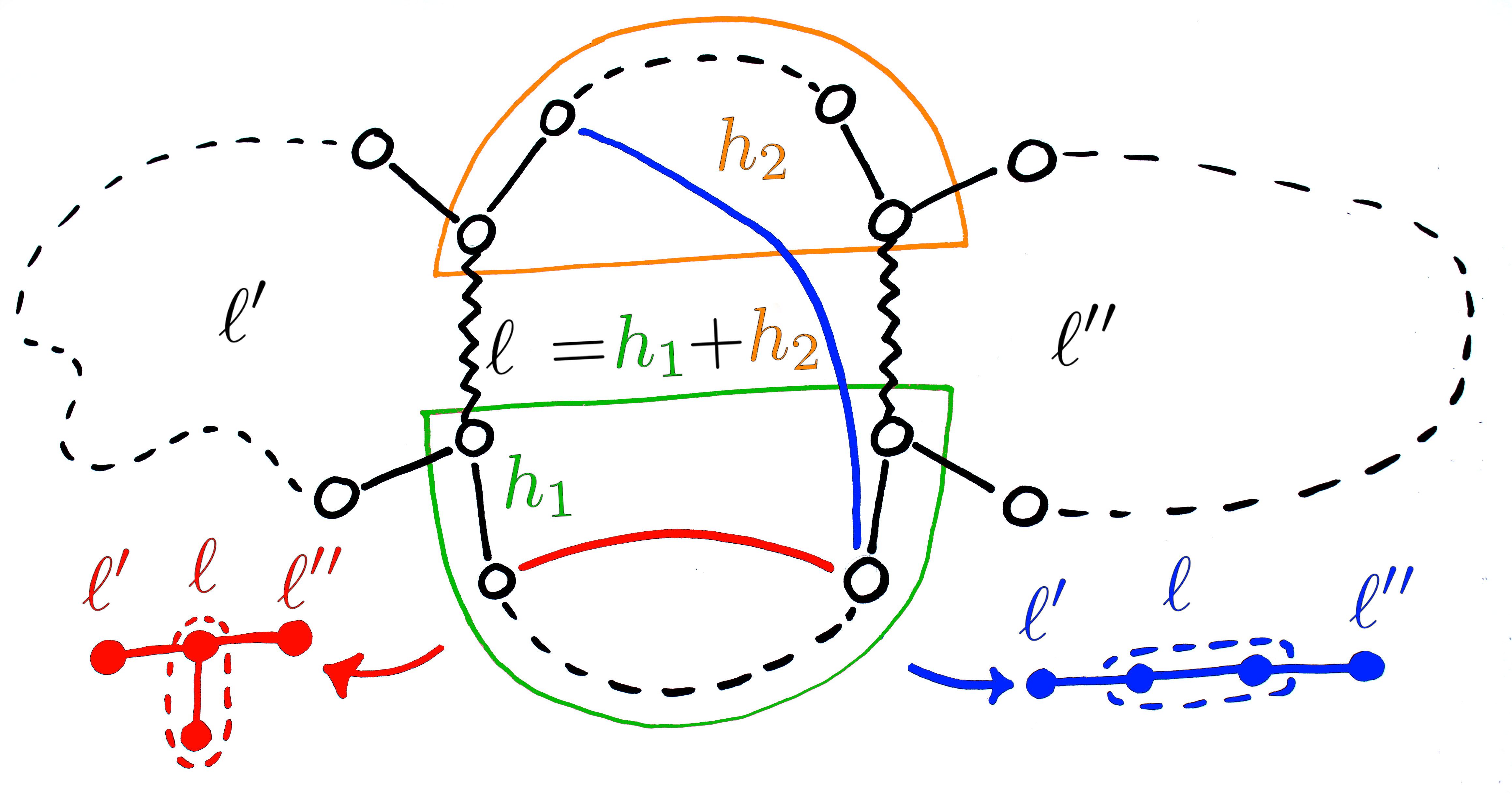}
        \caption{(color online) \textbf{Diameter-increasing vs. diameter-conserving link addition.} Example illustrating three cycles of length $\ell$, $\ell `$ and $\ell `'$ along the diameter graph with the dashed lines signifying the rest of the network. Adding a link (red) parallel to the diameter path leaves the diameter constant and creates a branch. Adding a link (blue) transverse to the path (and thus  parallel to the edges indicated by wiggled lines) increases the diameter by one. Out of the $\ell (\ell-3)/2$ potential links to add, $h_1 h_2 -2$ may add transversely.}
        \label{Fig:p}
\end{figure}

We now derive a recurrence relation for the average diameter $\evdel{D_t}$ to then estimate how the final diameter scales with the chain length. Let $\evdel{V_t}$ be the expected number of cycles on the diameter graph and let $\evdel{E_t}$ be the number of end cycles (degree-one vertices of the dual) on any diameter path, as shown in Fig.\,\ref{dual}.
The average diameter $\evdel{D_t}$ evolves with time in three different ways. First, if a new link divides a cycle that is not part of the diameter graph, the diameter $\evdel{D_t}$ stays unchanged. Second, if  a new link divides an end cycle of the diameter graph (Fig.\,\ref{dual}), which in mean field approximation  occurs with probability $\evdel{E_t}/t$,  $\evdel{D_t}$ grows by one. Finally, if in the diameter graph a new link divides a cycle that is not an end, which analogously occurs with probability $(\evdel{V_t}-\evdel{E_t})/t$, $\evdel{D_t}$ grows by one if the splitting is transverse to a diameter path, which in turn occurs with some probability $\pp_t^+$; otherwise, if the splitting is parallel to the diameter path, $\evdel{D_t}$ also remains unchanged, compare Fig.~\ref{Fig:p}.
We thus obtain the recurrence relation
\begin{equation}
 \evdel{D_{t+1}}=\evdel{D_t}+\frac{1}{t}\Big(\evdel{E_t}+(\evdel{V_t}-\evdel{E_t}) \pp_t^+\Big)
 \label{eq.d}
\end{equation}
for the expectation value of the diameter. It remains to estimate  $\pp_t^+$, $\evdel{E_t}$ and $\evdel{V_t}$ and then to iterate the recurrence relation in time to obtain the diameter of the final network.

\textit{Approximating $\pp_t^+$.}
To find $\pp_t^+$, we first compute the probability $\pp_t(D_t\textsf{ increases}|\ell)$ of the diameter increasing given that a link adds in a cycle of length $\ell$ on the diameter path \footnote{We assume that each cycle has two neighbouring cycles along the diameter. This assumption holds except if the cycle is the base of two branches of identical length, which becomes more and more unlikely with increasing $N$.}. There are two ways such a link can add, see Fig.~\ref{Fig:p}.
If adding a link splits splits the cycle parallel to the diameter path, the newly created cycle becomes a side arm of the path, leaving the diameter unchanged.
Alternatively, if the new link splits the cycle transversally to the direction of the path the diameter extends by one.  Let $h_1$ and $h_2(=\ell-h_1)$ be the numbers of units in the two fractions transversal to the diameter path (Fig.~\ref{Fig:p}). Increasing the diameter thus requires to connect one of the $h_1$  units to one of the other $h_2$ units. Then $\pp(D_t  \textsf{ increases} |\ell, h_1) = 2\frac{h_1(\ell-h_1)-2}{\ell(\ell-3)}$ , because there are $\ell(\ell-3)/2$  ways of connecting any two units in the cycle and  $h_1(\ell-h_1)$ ways of forming a transversal connection, the term ``$-2$'' taking care of the two links that already exist between the two fractions of the original cycle. 
As every splitting of the cycle into two parts is equally likely for part sizes $h_1\in \{1,\ldots,\ell-1\}$, we find 
\begin{equation}
 \pp_t(D_t \textsf{ increases} |\ell)=\frac{1}{\ell-1}\sum_{h_1=1}^{\ell-1}2\frac{h_1(\ell-h_1)-2}{\ell(\ell-3)}=\frac{4}{3 \ell}+\frac{1}{3}.
 \label{eq.p+}
\end{equation}
Finally, the probability $\pp_t(\ell)$ of picking a cycle of length $\ell$ on the diameter path and depends on the entire past history and cannot be rigorously derived. 
We thus approximate $\pp_t^+=\sum_{\ell=4}^{N} \pp_t(\textsf{D increases}|\ell) \pp_t(\ell) \approx \pp_t(\textsf{D increases}|\left< \ell_t \right>)$ by its rigorous lower bound given by Jensen's inequality.

We take the desired expected cycle length for sufficiently small times $t$ to be its average length $\left< \ell_t \right> = \frac{N+2(t-1)}{t}$ of all cycles at time $t$. As no links can be added to cycles of less then $\ell=4$ units, we take $\left< \ell_t \right> =4$ once the previous average reaches that value from above,  $\frac{N+2(t-1)}{t}\leq 4$, i.e., for $t\geq N/2$ for sufficiently large $N$, yielding
\begin{equation}
 \pp_t^+\approx \begin{cases} 
      \frac{4 t}{3(N+2(t-1))}+\frac{1}{3} & \textrm{for }t\leq N/2 , \\
      \frac{2}{3} &\textrm{for } t > N/2 . 
   \end{cases}
 \label{eq.p}
\end{equation}
We now approximate the detailed dynamics (\ref{eq.p}), by its time average, $\pp_t^+ \approx \overline{\pp^+}= (N-2)^{-1}\sum_{t=1}^{N-2}\pp_t^+\approx 0.602$.

 \textit{Approximating $\evdel{V_t}$ }
Next we estimate the average number of cycles in the diameter graph
\begin{equation}
 \evdel{V_t}=\evdel{D_t}+ \sum_b V_b(t) \pp_b(t)
\end{equation}
given by two contributions, the average diameter itself and the summed sizes $V_b(t)$ of all side branches $b$ of an arbitrary but fixed diameter path, weighted with the probability  $\pp_b$ that branch $b$ creates an alternative diameter path overlapping with the original. As longer side chains are exponentially suppressed, the second term is negligible for the scaling in the limit of large $N$ (see Supplemental Material for more details).

\textit{Iterated recurrence and scaling law.}
This suggests that $\evdel{D_t}$ and $\evdel{V_t}$ scale the same and therefore $\evdel{E_t}$ can be neglected in Eq.\,\ref{eq.d} without changing the scaling behaviour.\\
With $\evdel{V_t} \approx \evdel{D_t}$, the recurrence (\ref{eq.d}) becomes
\begin{equation}
 \evdel{D_{t+1}}\approx \evdel{D_t}+  \frac{\overline{\pp^+}}{t} \evdel{D_t} .
 \label{eq.d2}
\end{equation}
The solution through the initial condition $\evdel{D_{2}}=1$ is $\evdel{D_{t}}=2 \Gamma(\overline{\pp^+}+t)/ (\Gamma(\overline{\pp^+}+1)\Gamma(t)) \approx 2/\Gamma(\overline{\pp^+} +1) t^{\overline{\pp^+}}$,
where $\Gamma(.)$ is the Gamma function. In the limit of large $t=N-2$, a power law with specified exponent results, 
\begin{equation}
 \evdel{D_{N}}\sim  N^{\overline{\pp^+}}\quad \Rightarrow \quad \nu_{\textsf{theory}}=\overline{\pp^+} .
 \label{eqn:scalingtheory}
\end{equation} 
As found above already through direct numerical simulations, the scaling law now also obtained analytically is consistent with experimentally obtained law (\ref{eq:RG}) for proteins, with scaling exponent between the set upper and lower bounds (\ref{eq:exponentBounds}), compare Figs.\,\ref{num}b with \ref{num}a. Interestingly, the generally concave form of the dynamics of $\pp_t^+$, (see SM), indicates that any estimate of the time average $\nu_{\textsf{theory}}=\overline{\pp^+}$ must lie within an interval $\nu_{\textsf{theory}} \in [\nu_{\textsf{min}},\nu_{\textsf{max}}]$, where $\nu_{\textsf{max}}<2/3$ and 
$\nu_{\textsf{min}}>1/2$. Thus even without the approximation of the dynamics (\ref{eq.p}), an algebraic scaling is guaranteed and its exponent is above that for space filling aggregates, $\nu_{\textsf{theory}}>\nu_\textsf{SF}$.

The scaling law intrinsically results from the geometric constraints: without such constraints the process analyzed above exactly reduces to the formation of Watts-Strogatz small-world networks with new links randomly adding to a circular graph \cite{Watts1998,Newman2001,Grabow2012}; for sufficiently many links, the diameter of such networks exhibits logarithmic scaling that is thus inconsistent with the algebraic scaling we found. Roughly speaking, due to the geometric constraints, any new link between two units drastically increases the probability of creating further links in these units' respective neighborhoods. As a consequence, the structures cannot be arbitrarily compact. Our numerical results as well as analytic derivations above indicate that the spatial extent  is modified qualitatively, changing a logarithmic to an algebraic scaling law.

\textit{Conclusion and outlook.} Taken together, we uncovered an algebraic scaling law for network formation processes under geometric constraints. We have analyzed a spatial network formation model by mapping  geometric constraints in space to purely graph-theoretical constraints on the topological changes of a network. Direct numerical simulations as well as analytic mean field calculations strongly indicate a scaling law with the graph diameter growing algebraically with system size, representing spatially self-similar (`fractal') networks. This algebraic law scaling is largely independent of the details of the model setup and clearly induced by geometric constraints. Even without the time-averaging approximation of the dynamics (\ref{eq.p}) an algebraic scaling is guaranteed, exhibiting an exponent larger than that of a space filling aggregate, $\nu>\nu_{\textsf{SF}}$, thus indicating self-similar features. Both the algebraic scaling \emph{per se} and its exponent are consistent with experimentally observed scaling of protein tertiary structures in real space \cite{Danielsson2010,Hofmann2012,Hong2013}. More generally, our results may suggest that geometric constraints generically induce algebraic (rather than logarithmic) scaling laws of networks forming in space.

%\bibliographystyle{unsrt}
% \bibliographystyle{apsrev4-1}
% 
% \bibliography{proteins}

%

\end{document}